\documentclass{aa}  
\usepackage{graphicx}
\usepackage{txfonts}
\newcommand{\subO}{_{\rm o}}
\newcommand{\subS}{_{\rm s}}
\begin{document} 

\title{A simple approach to CO cooling in molecular clouds}
\author{A. P. Whitworth, S. E. Jaffa}
\institute{School of Physics and Astronomy, Cardiff University, Cardiff CF24 3AA, UK\\
\email{ant@astro.cf.ac.uk}}
\date{Received ; accepted}

\abstract
{Carbon monoxide plays an important role in interstellar molecular clouds, both as a coolant, and as a diagnostic molecule. However, a proper evaluation of the cooling rate due to CO requires a determination of the populations of many levels, the spontaneous and stimulated radiative de-excitation rates between these levels, and the transfer of the emitted multi-line radiation; additionally, this must be done for three isotopologues. It would be useful to have a simple analytic formulation that avoided these complications and the associated computational overhead; this could then be used in situations where CO plays an important role as a coolant, but the details of this role are not the main concern. We derive such a formulation here, by first considering the two asymptotic forms that obtain in the limits of (a) low volume-density and optical depth, and (b) high volume-density and optical depth. These forms are then combined in such a way as to fit the detailed numerical results from Goldsmith \& Langer (1978; hereafter GL78). The GL78 results cover low temperatures, and a range of physical conditions where the interplay of thermal and sub-thermal excitation, optical-depth effects, and the contributions from rare isotopologues, are all important. The fit is obtained using the Metropolis-Hastings method, and reproduces the results of GL78 well. It is a purely local and analytic function of state --- specifically a function of the density, $\rho$, isothermal sound speed, $a$, CO abundance, $X_{_{\rm CO}}$, and velocity divergence, $\nabla\cdot{\boldsymbol\upsilon}$. As an illustration of its use, we consider the cooling layer following a slow steady non-magnetic planar J-shock. We show that, in this idealised configuration, if the post-shock cooling is dominated by CO and its isotopologues, the thickness of the post-shock cooling layer is very small and approximately independent of the pre-shock velocity, $\upsilon\subO$, or pre-shock isothermal sound speed, $a\subO$.}


\maketitle

\section{Introduction}

Carbon monoxide, CO, is a critical molecule in the physics of molecular clouds and star formation. First, it is believed to be the next most abundant molecule in the Universe after molecular hydrogen, H$_2$. Second, as compared with H$_2$, it has a relatively high moment of inertia, and a permanent dipole moment, so it emits readily at the low temperatures found in molecular clouds, whereas H$_2$ does not; therefore CO and its isotopologues are often used to trace the structure and dynamics of molecular clouds \citep[e.g.][]{Romaetal2016}. Third, because it emits readily, it plays an important role in the thermal balance of molecular clouds, helping to maintain their pervasive low temperatures. However, the physics underlying the formation/destruction of CO, and the net line emission from CO, is complicated.

The pathways to CO formation are very diverse and uncertain; they depend on the formation of several other species, on reaction rates that are only known approximately, and on the agency of cosmic rays to produce molecular ions. Likewise, the destruction of CO is difficult to model, largely because it involves line radiation; as a consequence, evaluating self-shielding is difficult, and is made more difficult because self-shielding has to compete with some lines also being blocked by H$_2$, with dust attenuation, and, at high densities, with CO freeze-out. A variety of schemes has been proposed to estimate the abundance of CO without considering all these details \citep[e.g.][]{NelsLang1997,NelsLang1999}, and indeed these approximate schemes are the ones used in many interstellar chemistry codes \citep[e.g.][]{GlovClar2012b}. In the solar vicinity, it seems that gas-phase CO is usually the dominant form of carbon at volume densities in the range $3\times 10^2\,{\rm H}_2\,{\rm cm}^{-3}\la n_{_{\rm H_2}}\la 3\times 10^5\,{\rm H}_2\,{\rm cm}^{-3}$ and column-densities $N_{_{\rm H_2}}\ga 10^{21}\,{\rm cm}^{-2}$. At lower volume- and column-densities, CO gives way to atomic and ionic carbon (C$^{\rm o}$ and C$^+$), due to the slow rate of the two-body reactions leading to CO formation, and the strong ultraviolet irradiation which rapidly destroys CO. At higher volume-densities, CO appears to freeze out onto dust; parenthetically, CO's role as a coolant also becomes less important at these higher densities, because the gas starts to couple thermally to the dust, and hence to cool by continuum emission from dust.

Line emission from CO is complicated by the fact that, in the range of volume- and column-density where CO tends to be abundant, transfer of population between the different levels on the rotational ladder of CO involves a balance between collisional and radiative excitation and de-excitation, and the details of this balance shift with changing volume-density, column-density, temperature and velocity dispersion. CO line emission is seldom in either of the asymptotic limits of very low volume- and column-density (hence low line-centre optical depth), or very high volume- and column-density (hence high line-centre optical depth), where its collective line emission can be described by simple algebraic equations \citep[e.g.][]{GoldKwan1974}. 

Here we derive an approximate analytic formulation for the net cooling rate from CO and its isotopologues. The formulation is simple in the sense that it is a purely local function of state, depending only on four parameters: (i) the mass-density, $\rho$, or equivalently the volume-density of molecular hydrogen, $n_{_{\rm H_2}}$; (ii) the isothermal sound speed, $a$, or equivalently the gas-kinetic temperature, $T$; (iii) the relative abundance of CO, $X_{_{\rm CO}}\equiv n_{_{\rm CO}}/n_{_{\rm H_2}}$; and (iv) the velocity divergence, $\nabla\cdot{\boldsymbol\upsilon}$. The formulation is obtained by first deriving the simple algebraic equations that describe the dependence of the CO cooling rate on density, temperature, CO abundance and velocity divergence, in the asymptotic limit of low volume-density and optical depth, and in the asymptotic limit of high volume-density and optical depth; hereafter we refer to these equations as the asymptotic forms. Their derivation is based on fundamental physical arguments, and forms the subject of Section \ref{SEC:COcoolingTheory}. Then, in Section \ref{SEC:COcoolingCalibration} the coefficients in front of the asymptotic forms, and the variation between them, are fit by reproducing the detailed results of GL78, using an ad hoc mathematical function. In Section \ref{SEC:PSC} we illustrate the application of the approximate analytic formulation by using it to evaluate the thickness of the post-shock cooling layer behind a low-velocity steady J-shock in a non-magnetic molecular cloud. In Section \ref{SEC:CONC} we summarise our conclusions.

For mathematical convenience, we use the mass-density, $\rho$ and the isothermal sound speed, $a\!=\!(k_{_{\rm B}}T/{\bar m})^{1/2}\;$ (in place of the number-density of molecular hydrogen, $n_{_{\rm H_2}}$, and the gas-kinetic temperature $T)\;$ in much of the analysis. Here $k_{_{\rm B}}$ is Boltzmann's constant and ${\bar m}$ is the mean gas-particle mass. For the purpose of illustration, we use a reference temperature $T\subO\!=\!10\,{\rm K}$ and a mean gas-particle mass ${\bar m}\!=\!3.97\times 10^{-24}\,{\rm g}$ (appropriate for molecular gas with elemental composition $X=0.70,\;Y=0.28,\;Z=0.02$);\footnote{In accordance with standard practice, $X,\,Y,\,Z$ are here the abundances {\it by mass} of hydrogen, helium and other elements (the `heavies'), whereas $X_{_{\rm CO}}\!\equiv\!n_{_{\rm CO}}/n_{_{\rm H_2}}$ is the abundance of CO relative to H$_{_2}$ {\it by number}. For simplicity, we assume that all the hydrogen is molecular.} hence the reference isothermal sound speed is $a\subO\!=\!0.187\,{\rm km}\,{\rm s}^{-1}$. On the assumption that virtually all the hydrogen is in the form of H$_2$, we also define the mean mass per hydrogen molecule, ${\bar m}_{_{\rm H_2}}=\rho/n_{_{\rm H_2}}\!=\!4.77\times 10^{-24}\,{\rm g}$, and hence $\rho=n_{_{\rm H_2}}{\bar m}_{_{\rm H_2}}$.

\section{The CO cooling rate: asymptotic theory}\label{SEC:COcoolingTheory}

In this section we derive, using basic physical arguments, the dependence of the CO line cooling rate on the density, on the isothermal sound speed (or equivalently the temperature), on the gas-phase CO abundance, and on the local velocity divergence. We treat the two asymptotic limits of (i) very low volume-density and optical depth (hereafter the `{\sc lo}' limit), and (ii) very high volume-density and optical depth (hereafter the `{\sc hi}' limit). We do not derive the absolute physical value of the cooling rate in these limits, but only the dependence on density, isothermal sound speed, abundance and velocity divergence. The coefficients converting these dependences into actual physical values are obtained in the following section (Section \ref{SEC:COcoolingCalibration}) by comparing the theoretical predictions with the detailed results of GL78.

\subsection{CO cooling in the {\sc LO} limit}

For a linear molecule with moment of inertia $I_{_{\rm MOL}}$, the rotational levels are characterised by quantum number $J$, and have energy $\,E_{_J}=J(J+1)\hbar^2/2I_{_{\rm MOL}};\,$ therefore, if the gas-kinetic temperature is $T$, they are significantly excited up to 
\begin{eqnarray}\label{EQN:JMAX}
J_{_{\rm MAX}}&\simeq&f_{_{\rm EX}}\frac{(2I_{_{\rm MOL}}k_{_{\rm B}}T)^{1/2}}{\hbar}\,,
\end{eqnarray}
where $f_{_{\rm EX}}$ is a factor greater than, but of order, unity. Radiative de-excitations from level $J$ to level $J\!-\!1$ release photons with energy $\Delta E_{_J}=J\hbar^2/I_{_{\rm MOL}}$. Therefore the mean energy of emitted photons is
\begin{eqnarray}\nonumber
\overline{\Delta E}&\simeq&\int\limits_{J=0}^{J=J_{_{\rm MAX}}}\Delta E_{_J}\,(2J+1)\,dJ\left\{\int\limits_{J=0}^{J=J_{_{\rm MAX}}}(2J+1)\,dJ\right\}^{-1}\\\label{EQN:DeltaE}
&\simeq&f_{_{\rm EX}}\,\hbar\,\left(\frac{8k_{_{\rm B}}T}{9I_{_{\rm MOL}}}\right)^{1/2}.
\end{eqnarray}
Here, we have replaced a sum over levels with an integral. This is reasonable for higher temperatures, $T\ga 20\,{\rm K}$, but may introduce a small systematic error at lower temperatures where only a few levels are involved.

At sufficiently low {\it volume-density}, i.e. well below the critical density, the level populations are not thermalised. Most molecules sit in their ground state, and most collisional excitations are followed by radiative de-excitations. For example, at $T\sim10\,{\rm K}$, the first rotationally excited level of CO,  $J=1$, has a critical density of order $10^3\,{\rm H_2\,cm}^{-3}$; for higher-$J$ levels, higher temperatures and densities are required before they are thermalised. At sufficiently low {\it optical depth} (i.e. sufficiently low column-density and/or high velocity divergence), the lines are optically thin, so most of the emitted photons escape directly; consequently the thermal kinetic energy that caused the initial excitation is lost, and the gas is cooled. For example, at $T\sim 10\,{\rm K}$ and with purely thermal velocity dispersion, i.e. no turbulence and no velocity gradient, the first rotational transition of CO ($J=1\rightarrow 0$) becomes optically thick for column-densities $N_{_{\rm H_2}}\ga 10^{19}\,{\rm H_2\,cm}^{-2}$; this limit is increased if there is turbulence and/or a velocity gradient delivering extra velocity dispersion. The rate of collisional excitation per unit volume depends on the product of the number-densities of the molecule and the exciting collider (here presumed to be H$_2$, see below) and their relative speed, so it is approximately proportional to $n_{_{\rm MOL}}n_{_{{\rm H}_2}}T^{1/2}$. The number of lines that are excited is approximately proportional to $T^{1/2}$ (see Eqn. \ref{EQN:JMAX}). And the mean energy of the emitted photons is approximately proportional to $T^{1/2}$ (see Eqn. \ref{EQN:DeltaE}). Therefore, in the limit of very low volume-density and optical depth, the cooling rate per unit volume can be approximated by
\begin{eqnarray}\label{EQN:Lambda_MOL>.LO}
\Lambda_{_{\rm MOL.LO}}&\propto& n_{_{\rm MOL}}n_{_{{\rm H}_2}}T^{3/2}\;\,\propto\;\,X_{_{\rm MOL}}\,\rho^2a^3\,.
\end{eqnarray}
In treating cooling by CO, we presume that H$_2$ is the dominant agent of collisional excitation, since, in regions where there is CO, H$_2$ is likely to be by far the most abundant species; GL78 make the same assumption. 

On the basis of Eqn. (\ref{EQN:Lambda_MOL>.LO}), we posit that, in the {\sc lo} limit, the cooling rate due to CO is given by
\begin{eqnarray}\label{EQN:LambdaCOLO.1}
\Lambda_{_{\rm CO.LO}}&\simeq&\lambda_{_{\rm CO.LO}}\,X_{_{\rm CO}}\,\rho^2a^3\,.
\end{eqnarray}
The coefficient $\lambda_{_{\rm CO.LO}}$ will be estimated in Section \ref{SEC:COcoolingCalibration}, but the above dependence of $\Lambda_{_{\rm CO.LO}}$ on $X_{_{\rm CO}}$, $\rho$ and $a$ is now fixed.

\begin{figure*}
\includegraphics[width=130mm,angle=270]{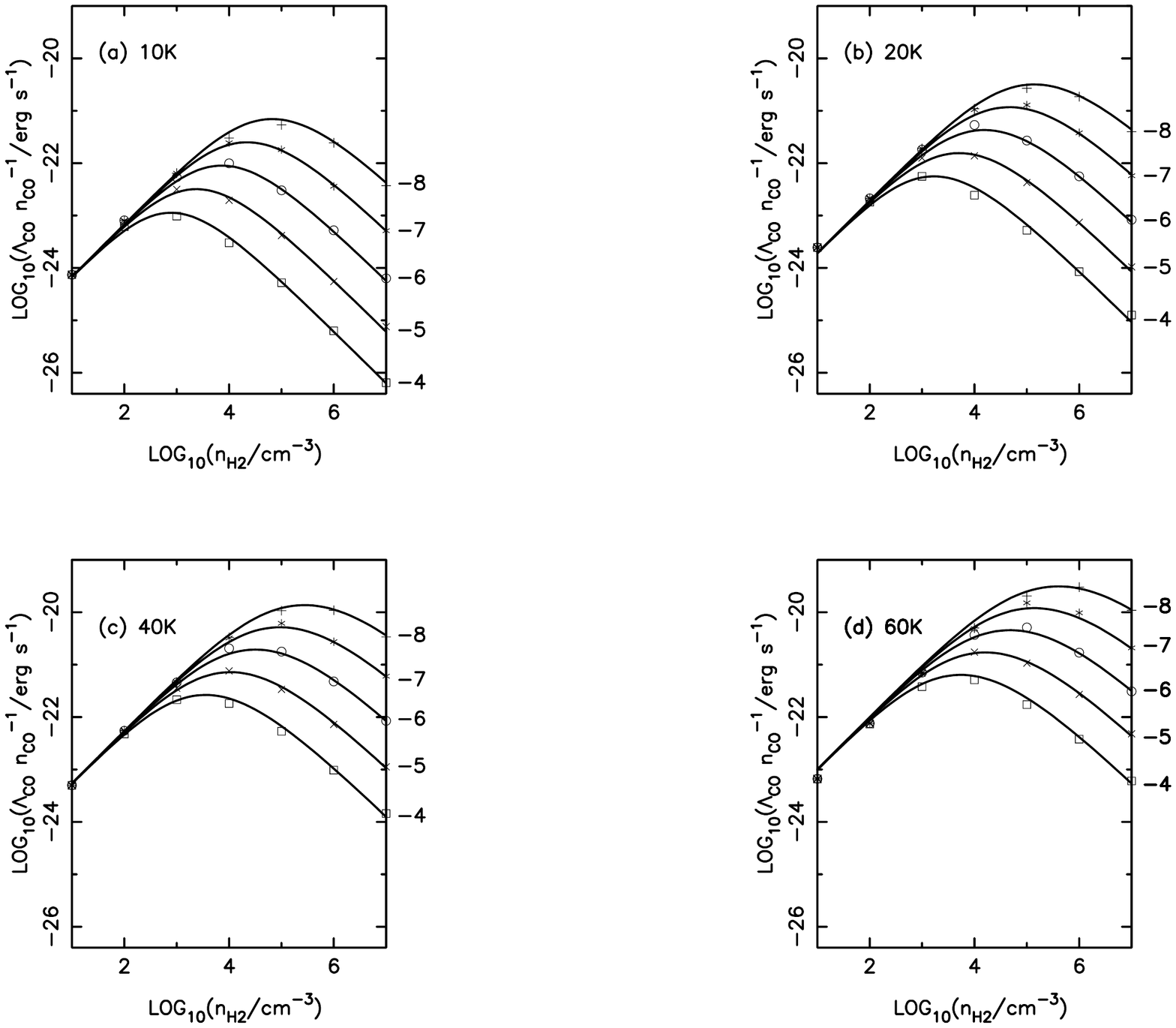}
\caption{The continuous curves give the total CO cooling rate, as predicted by the approximate analytic formulation derived here (Eqns. \ref{EQN:Lambda_COLO.2} through \ref{EQN:Lambda_COTOT.2}). The different panels correspond to (a) $T\!=\!10\,{\rm K}$; (b) $T\!=\!20\,{\rm K}$; (c) $T\!=\!40\,{\rm K}$; and (d) $T\!=\!60\,{\rm K}$ (these temperatures are given in the top lefthand corner of each panel). The different curves represent different values of $\log_{_{10}}\left\{X_{_{\rm CO}}(|\nabla\cdot{\boldsymbol\upsilon}|/3\,{\rm km}\,{\rm s}^{-1}\,{\rm pc}^{-1})^{-1}\right\}=-8,\;-7,\;-6,\;-5\;{\rm and}\;-4$ (these values are given down the righthand margin of each panel). The rates obtained by the detailed computations of GL78 are given at discrete values of the density, $\log_{_{10}}\left\{n_{_{\rm H_2}}/{\rm cm}^{-3}\right\}=1,\;2,\;3,\;4,\;5,\;6\;{\rm and}\;7$; plus signs, stars, open circles, crosses and open squares correspond respectively to $\log_{_{10}}\left\{X_{_{\rm CO}}(|\nabla\cdot{\boldsymbol\upsilon}|/3\,{\rm km}\,{\rm s}^{-1}\,{\rm pc}^{-1})^{-1}\right\}=-8,\;-7,\;-6,\;-5\;{\rm and}\;-4$.}
\label{FIG:GoldLangFig2}
\end{figure*}

\subsection{CO cooling in the {\sc HI} limit}

At sufficiently high {\it volume-density}, i.e. well above the critical density, the level populations are approximately thermalised, and radiative de-excitations account for only a small fraction of de-excitations. At sufficiently high {\it optical depth} (i.e. sufficiently high column-density and/or low velocity divergence), the intensity in the emission line cores is approximately equal to the Planck Function, and the net frequency width of all the excited lines is 
\begin{eqnarray}
\Delta\nu_{_{\rm NET}}&=&\int\limits_{J\!=\!0}^{J\!=\!J_{_{\rm MAX}}}\,\frac{\Delta E_{_J}}{h}\,\frac{f_{_{\rm LW}}\Delta \upsilon}{c}\,dJ\;\,\simeq\;\, f_{_{\rm EX}}^2\,\frac{k_{_{\rm B}}T}{h}\,\frac{f_{_{\rm LW}}\Delta \upsilon}{c}\,.
\end{eqnarray}
Here $\Delta \upsilon$ is the velocity dispersion along the line of sight and $f_{_{\rm LW}}$ is the factor by which the effective width of an optically thick line with central frequency $\nu\subO$ differs from\footnote{If the equivalent width of an optically thick emission line falls on the flat portion of the curve of growth, this factor is approximately constant, and that is what we assume here. If the equivalent width falls on the square-root portion of the curve of growth, $f_{_{\rm LW}}$ depends on the column-density (as $N_{_{\rm MOL}}^{1/2}$). However, this requires extremely large column-densities. By the time such high column-densities are reached, molecules like CO are likely to have frozen out --- and, even if they have not, the density is so high that molecular-line cooling has given way to dust cooling.} $\,\nu\subO \Delta \upsilon/c$. The effective width of the Planck spectrum is $\Delta\nu_{_{\rm BB}}=f_{_{\rm BB}}k_{_{\rm B}}T/h$ where $f_{_{\rm BB}}=14.4$, and so the total cooling rate per unit volume in this limit can be approximated by
\begin{eqnarray}\nonumber
\Lambda_{_{\rm MOL.HI}}&\simeq&\frac{6\sigma_{_{\rm SB}}T^4}{D}\,\frac{\Delta\nu_{_{\rm NET}}}{\Delta\nu_{_{\rm BB}}}\\\label{EQN:LambdaMolHI.1}
&\simeq&\frac{6\sigma_{_{\rm SB}}T^4}{c}\,\frac{\Delta \upsilon}{D}\,\frac{f_{_{\rm EX}}^2f_{_{\rm LW}}}{f_{_{\rm BB}}}\\\label{EQN:LambdaMolHI.2}
&\propto&a^8\,|\nabla\cdot{\boldsymbol\upsilon}|
\end{eqnarray}
In deriving Eqn. (\ref{EQN:LambdaMolHI.1}), we have adopted the most generic geometry we could imagine, viz. a spherical cloud expanding or collapsing homologously, so that $\Delta \upsilon/D\equiv |\nabla\cdot{\boldsymbol\upsilon}|/3$; GL78 adopt the same configuration. However, in obtaining Eqn. (\ref{EQN:LambdaMolHI.2}), this choice becomes immaterial, since we only retain the dependence on $a$ and $\nabla\cdot{\boldsymbol\upsilon}$. The effect of geometry is encapsulated entirely in the $\nabla\cdot{\boldsymbol\upsilon}$ term, which measures how far, in different directions on the sky, line-photons have to travel before the Doppler-shift has separated their frequency from the frequencies that can be absorbed by the molecules they are passing.

In effect we are adopting the large velocity gradient (LVG) approximation to treat optically thick cooling radiation. The LVG approximation was introduced by \citet{SobolevV1960}, and developed further by \citet{CastorJI1970} and \citet{LucyLB1971}, in the context of stellar outflows, but it can also be applied to the cooling of molecular clouds \citep[e.g.][]{GoldKwan1974}. Basically it assumes that line radiation only experiences self-absorption in a local region whose linear size is of order $\sigma\,|\nabla\cdot{\boldsymbol\upsilon}|^{-1}$, where $\sigma$ is the local velocity dispersion (thermal plus turbulent) and ${\boldsymbol\upsilon}$ is the local bulk velocity. Any material outside this region is moving at a sufficiently different velocity (bulk plus or minus the dispersion) from the emitting region that it cannot absorb its line radiation. Strictly, the LVG approximation requires that the large-scale spatial variation of the bulk velocity is monotonic along any line of sight.

On the basis of Eqn. (\ref{EQN:LambdaMolHI.2}), we posit that, in the {\sc hi} limit, the CO cooling rate is given by
\begin{eqnarray}\label{EQN:LambdaCOHI.1}
\Lambda_{_{\rm CO.HI}}&\simeq&\lambda_{_{\rm CO.HI}}\,a^8\,|\nabla\cdot{\boldsymbol\upsilon}|\,;
\end{eqnarray}
the coefficient $\lambda_{_{\rm CO.HI}}$ will be estimated in Section \ref{SEC:COcoolingCalibration}, but the above dependence of $\Lambda_{_{\rm CO.HI}}$ on $a$ and $|\nabla\cdot{\boldsymbol\upsilon}|$ is now fixed --- and $\Lambda_{_{\rm CO.HI}}$ is independent of $\rho$ and $X_{_{\rm CO}}$.

We note parenthetically that the only parameters in Eqn. (\ref{EQN:LambdaMolHI.1}) that depend on the specific molecule are $f_{_{\rm EX}}$ and $f_{_{\rm LW}}$, and that this dependence is in general rather weak, so Eqn. (\ref{EQN:LambdaMolHI.1}) is approximately valid for many other linear molecules in this high volume-density, high optical depth limit. Thus, for example, the results presented by GL78 for cooling by O$_2$ are very similar to those for CO (at the same density and temperature).

\subsection{In between the {\sc LO} and {\sc HI} limits}

In the next section (Section \ref{SEC:COcoolingCalibration}) we show how Eqns. (\ref{EQN:LambdaCOLO.1}) and (\ref{EQN:LambdaCOHI.1}) can be combined to obtain an approximate analytic formulation for the total CO cooling rate in the intermediate regime (between the {\sc lo} and {\sc hi} limits) using
\begin{eqnarray}\label{EQN:LambdaCOTOT.1}
\Lambda_{_{\rm CO.TOT}}&=&\left\{\Lambda_{_{\rm CO.LO}}^{-1/\beta}+\Lambda_{_{\rm CO.HI}}^{-1/\beta}\right\}^{-\beta}\,,\end{eqnarray}
with $\beta=\beta(\rho,a)$.

\section{The CO cooling rate: calibration}\label{SEC:COcoolingCalibration}

In the preceding section (Section \ref{SEC:COcoolingTheory}) we have obtained expressions for the CO cooling rate (in the {\sc lo} and {\sc hi} limits) in terms of the mass-density, $\rho$, the isothermal sound speed, $a$, the abundance of CO, $X_{_{\rm CO}}$, and the velocity divergence, $\nabla\cdot{\boldsymbol\upsilon}$. This is because these are the most convenient variables to use when treating the equations of hydrodynamics. Here, we replace $\rho$ with the density of molecular hydrogen $n_{_{\rm H_2}}$, and $a$ with the temperature, $T$, since these are the variables used by GL78. It is straightforward to switch between the two, using $n_{_{\rm H_2}}=\rho/{\bar m}_{_{\rm H_2}}$ and $T={\bar m}a^2/k_{_{\rm B}}$.

\subsection{The detailed results of \citet{GoldLang1978}}

GL78 consider a uniform-density spherical cloud with a linear radial velocity field, and determine the net cooling rate, $\Lambda_{_{\rm CO.TOT}}$, from  CO and its isotopologues. To do this, they solve in detail (i) the equations of statistical equilibrium, to determine CO level populations; then (ii) the equations of radiative emission (spontaneous and stimulated), to determine the net CO emissivity; finally (iii) they use an escape probability in lieu of solving the equation of radiative transfer. In their Fig. 2, which is the input we use to calibrate our approximate analytic formulation, they plot the quantity $\log_{_{10}}\{\Lambda_{_{\rm CO.TOT}}/n_{_{\rm CO}}\}$ against $\log_{_{10}}\{n_{_{\rm H_2}}/{\rm cm}^{-3}\}$, for different values of $T$ and different values of \footnote{The factor of 3 here, and in subsequent expressions, derives from the fact that we have replaced Goldsmith and Langer's $d\upsilon/dr$ with $\nabla\cdot{\boldsymbol\upsilon}=3\,d\upsilon/dr$.} $\,\log_{_{10}}\left\{X_{_{\rm CO}}\left(|\nabla\cdot{\boldsymbol\upsilon}|/3\,{\rm km}\,{\rm s}^{-1}\,{\rm pc}^{-1}\right)^{-1}\right\}$, specifically $T=10,\,20,\,40\;{\rm and}\;60\,{\rm K}$ and $\log_{_{10}}\left\{X_{_{\rm CO}}\left(|\nabla\cdot{\boldsymbol\upsilon}|/3\,{\rm km}\,{\rm s}^{-1}\,{\rm pc}^{-1}\right)^{-1}\right\}=-8,\,-7,\,-6,\,-5\;{\rm and}\;-4$. We note that $\Lambda_{_{\rm CO.TOT}}/n_{_{\rm CO}}$ is the cooling rate per CO molecule. It depends on $X_{_{\rm CO}}/|\nabla\cdot{\boldsymbol\upsilon}|$, because -- all other things being equal -- the optical depth increases with increasing $X_{_{\rm CO}}$ and decreases with increasing $|\nabla\cdot{\boldsymbol\upsilon}|$. In the {\sc lo} limit, where the emission is optically thin, $\Lambda_{_{\rm CO.TOT}}/n_{_{\rm CO}}$ is independent of $X_{_{\rm CO}}/|\nabla\cdot{\boldsymbol\upsilon}|$.

Given the complexity of the physics underlying the net CO cooling rate, and the consequent computational cost of calculating it properly, it would be useful to have an approximate analytic formulation that captured the dependence on local variables but avoided the associated computational cost. This formulation could then be used in situations where CO cooling played an important role, but was not the principal interest, as for example, contracting pre-stellar cores, or the accretion shock at the boundary of an assembling filament.

\subsection{The fitting procedure}

To fit the GL78 results analytically, we have read from their plots the values of $\log_{_{10}}\{\Lambda_{_{\rm CO.TOT}}/n_{_{\rm CO}}\}$ for each treated combination of $T$, i.e. $T=10,\,20,\,40\;{\rm and}\;60\,{\rm K}$, and $\log_{_{10}}\left\{X_{_{\rm CO}}\left(|\nabla\cdot{\boldsymbol\upsilon}|/3\,{\rm km}\,{\rm s}^{-1}\,{\rm pc}^{-1}\right)^{-1}\right\}$, i.e. $\log_{_{10}}\left\{X_{_{\rm CO}}\left(|\nabla\cdot{\boldsymbol\upsilon}|/3\,{\rm km}\,{\rm s}^{-1}\,{\rm pc}^{-1}\right)^{-1}\right\} =-8,\, -7,\, -6,\, -5\; {\rm and}\;  -4$, at the densities $\log_{_{10}}\{n_{_{\rm H_2}}/{\rm cm}^{-3}\}=1,\, 2,\, 3,\, 4,\, 5,\, 6,\, {\rm and}\, 7$, thus a total of 140 discrete values. We estimate that the uncertainty which derives from our reading these values off GL78's Fig. 2 by eye is $<0.03$ in $\log_{_{10}}\{\Lambda_{_{\rm CO.LO}}/n_{_{\rm CO}}\}$ i.e. at worst $\,\pm 3\%$, and less than the thickness of the lines on Fig. \ref{FIG:GoldLangFig2}.

We then use a Metropolis-Hastings Markov Chain Monte Carlo algorithm to find the five fitting parameters $(\lambda'_{_{\rm CO.LO}},\lambda'_{_{\rm CO.HI}},\beta'_{_{\rm O}},\beta'_{n_{\rm H_2}},\beta'_T)$ in the formulation
\begin{eqnarray}
\frac{\Lambda_{_{\rm CO.LO}}}{n_{_{\rm CO}}}&=&\lambda'_{_{\rm CO.LO}}\,\left(\!\frac{n_{_{\rm H_2}}}{{\rm cm}^{-3}}\!\right)\,\left(\!\frac{T}{\rm K}\!\right)^{\!3/2}\,,\\
\frac{\Lambda_{_{\rm CO.HI}}}{n_{_{\rm CO}}}&=&\lambda'_{_{\rm CO.LO}}\,\left(\frac{X_{_{\rm CO}}\,{\rm km}\,{\rm s}^{-1}\,{\rm pc}^{-1}}{|\nabla\cdot{\boldsymbol\upsilon}|}\right)^{-1}\left(\frac{n_{_{\rm H_2}}}{{\rm cm}^{-3}}\right)^{-1}\left(\frac{T}{\rm K}\right)^{\!4}\,,\\
\beta&=&\beta'\subO\,\left(\frac{n_{_{\rm H_2}}}{{\rm cm}^{-3}}\right)^{\beta'_{n_{\rm H_2}}}\,\left(\frac{T}{\rm K}\right)^{\beta'_T}\,,\\\label{EQN:Lambda_COTOT.2}
\frac{\Lambda_{_{\rm CO.TOT}}}{n_{_{\rm CO}}}&=&\left\{\left(\frac{\Lambda_{_{\rm CO.LO}}}{n_{_{\rm CO}}}\right)^{-1/\beta}+\left(\frac{\Lambda_{_{\rm CO.HI}}}{n_{_{\rm CO}}}\right)^{-1/\beta}\right\}^{-\beta}\,,
\end{eqnarray}
that give the best fit to these discrete values. 

We note that the exponents from Eqns. (\ref{EQN:LambdaCOLO.1}) and (\ref{EQN:LambdaCOHI.1}) are not being allowed to vary, they retain the values derived in Section \ref{SEC:COcoolingTheory}. Because we are here fitting $\Lambda_{_{\rm CO.TOT}}/n_{_{\rm CO}}$ (rather than $\Lambda_{_{\rm CO.TOT}}$), the $\rho^2$ term in Eqn. (\ref{EQN:LambdaCOLO.1}) has become $n_{_{\rm H_2}}^1$, and the $\rho^0$ term implicit in Eqn. (\ref{EQN:LambdaCOHI.1}) has become $n_{_{\rm H_2}}^{-1}$. The $a^3$ term in Eqn. (\ref{EQN:LambdaCOLO.1}) has become $T^{3/2}$ and the $a^8$ term in Eqn. (\ref{EQN:LambdaCOLO.1}) has become $T^4$.

\subsection{The best fit and its accuracy}

The best fit to the results from GL78 is obtained with the following values for the fitting parameters:
\begin{eqnarray}
\left.\begin{array}{lll}
\lambda'_{_{\rm CO.LO}}&=&2.16\times 10^{-27}\,{\rm erg}\,{\rm s}^{-1}\,, \\
\lambda'_{_{\rm CO.HI}}&=&2.21\times 10^{-28}\,{\rm erg}\,{\rm s}^{-1}\,, \\
\beta'\subO&=&1.23\,, \\
\beta'_{n_{H_2}}&=&0.0533\,, \\
\beta'_T&=&0.164\,.\\
\end{array}\right\}
\end{eqnarray}
In other words, the best fit is
\begin{eqnarray}\label{EQN:Lambda_COLO.2}
\frac{\Lambda_{_{\rm CO.LO}}}{n_{_{\rm CO}}}&=&\left[2.16\times 10^{-27}\,{\rm erg}\,{\rm s}^{-1}\right]\,\left(\!\frac{n_{_{\rm H_2}}}{{\rm cm}^{-3}}\!\right)\,\left(\!\frac{T}{\rm K}\!\right)^{\!3/2}\,;\\\nonumber
\frac{\Lambda_{_{\rm CO.HI}}}{n_{_{\rm CO}}}&=&\left[2.21\times 10^{-28}\,{\rm erg}\,{\rm s}^{-1}\right]\\\label{EQN:Lambda_COHI.2}
&&\hspace{0.6cm}\times\,\left(\frac{X_{_{\rm CO}}\,{\rm km}\,{\rm s}^{-1}\,{\rm pc}^{-1}}{|\nabla\cdot{\boldsymbol\upsilon}|}\right)^{-1}\left(\frac{n_{_{\rm H_2}}}{{\rm cm}^{-3}}\right)^{-1}\left(\frac{T}{\rm K}\right)^{\!4}\,;\\\label{EQN:beta.2}
\beta&=&1.23\,\left(\frac{n_{_{\rm H_2}}}{{\rm cm}^{-3}}\right)^{0.0533}\,\left(\frac{T}{\rm K}\right)^{0.164}\,;\\\label{EQN:Lambda_COTOT.2}
\frac{\Lambda_{_{\rm CO.TOT}}}{n_{_{\rm CO}}}&=&\left\{\left(\frac{\Lambda_{_{\rm CO.LO}}}{n_{_{\rm CO}}}\right)^{-1/\beta}+\left(\frac{\Lambda_{_{\rm CO.HI}}}{n_{_{\rm CO}}}\right)^{-1/\beta}\right\}^{-\beta}\,.
\end{eqnarray}

The continuous lines on Fig. \ref{FIG:GoldLangFig2} show the predictions of Eqns. (\ref{EQN:Lambda_COLO.2}) through (\ref{EQN:Lambda_COTOT.2}) for $T=10,\,20,\,40\;{\rm and}\;60\,{\rm K}$ (respectively panels a,b,c,d), and $\log_{_{10}}\left\{X_{_{\rm CO}}\left(|\nabla\cdot{\boldsymbol\upsilon}|/3\,{\rm km}\,{\rm s}^{-1}\,{\rm pc}^{-1}\right)^{-1}\right\}=-8,\,-7,\,-6,\,-5\;{\rm and}\;-4$ (separate lines labelled on the righthand margin of each panel). Like GL78 we limit the plots to densities in the range $1\leq\log_{_{10}}\{n_{_{\rm H_2}}/{\rm cm}^{-3}\}\leq 7$. The continuous curves should be compared with the discrete symbols, which give the values that we have read from Fig. 2 of GL78 and that we used in the Metropolis-Hastings fitting procedure.

In general, the agreement is good, with the magnitude of the fractional offset being everywhere $<\!0.18$ in $\log_{_{10}}\{\Lambda_{_{\rm CO.LO}}/n_{_{\rm CO}}\}\;\;\left(\mbox{i.e., at worst, }_{-34\%}^{+52\%}\right)$, and on average $\la\!0.07$ in $\log_{_{10}}\{\Lambda_{_{\rm CO.LO}}/n_{_{\rm CO}}\}\;\;\left(\mbox{i.e., typically, }_{-15\%}^{+18\%}\right)$. The fractional offset is worst in the intermediate regime. This is as expected, since the analytic forms that obtain in the asymptotic limits ({\sc lo} and {\sc hi}) are physically motivated, and therefore in some sense absolute. In contrast, the intermediate regime between these limits is being fit with an algebraic expression that has no physical motivation beyond the fact that it relaxes to the asymptotic forms in the corresponding limits; it is therefore unable to produce an exact fit. The error in reading the discrete values off GL78's Fig. 1 by eye ($\pm 3\%$) is much smaller than the fractional offsets cited above, and can therefore be ignored.

There are some additional uncertainties that should be noted. First, GL78 used the coefficients for collisional de-excitation of CO by H$_2$ computed by \citet{GreeThad1976}. More recent computations by \citet{Yangetal2010} have increased these coefficients somewhat, at the lowest temperatures, $T\la 20\,{\rm K}$. Therefore the GL78 cooling rates, and our approximate analytic formulation based on them, may be a little low at these low temperatures. Second, the device of replacing with integrals, summations over the discrete contributions from individual levels, will be most inaccurate at low temperatures where only a few levels are involved.

We note that the rarer isotopologues of CO make little contribution to the cooling in the {\sc lo} limit, but do contribute once the $^{12}$C$^{16}$O lines start to become optically thick, because the rarer isotopologues remain optically thin to much higher column-densities. They also make an important contribution in the {\sc hi} limit, by providing extra bandwidth at high column-densities; in fact, in the {\sc hi} limit, each isotopologue makes essentially the same contribution to the net cooling rate.

\subsection{Use and range of applicability of the approximate analytic formulation}

Given this level of accuracy, the approximate analytic formulation (i.e. Eqns. \ref{EQN:Lambda_COLO.2} through  \ref{EQN:Lambda_COTOT.2}) provides a convenient and computationally very inexpensive way to estimate the net CO cooling rate: convenient because it is a local function of state, and computationally inexpensive because it entails only a small number of arithmetic operations. It can therefore be used to greatly speed up numerical and/or analytic integrations of the energy equation in hydrodynamic or hydrostatic simulations, provided only that the conditions are within the range explored by GL78. 

The numerical results of GL78 cover the density range $10\,{\rm cm}^{-3}\leq n_{_{\rm H_2}}\la 10^7\,{\rm cm}^{-3}$, the temperature range $10\,{\rm K}\leq T\leq 60\,{\rm K}$, and the abundance/velocity-divergence range $-8\la \log_{_{10}}\left\{X_{_{\rm CO}}\left(|\nabla\cdot{\boldsymbol\upsilon}|/3\,{\rm km}\,{\rm s}^{-1}\,{\rm pc}^{-1}\right)^{-1}\right\}\la -4$, and therefore this is the range over which our approximate analytic formulation is most secure. Given the regularity of the curves in GL78, and their apparent self-similarity, it may be safe to extrapolate further. It is unlikely that gas-phase CO is the dominant form of CO outside the density range defined above, so we will only consider extrapolation of the temperature and abundance/velocity-divergence ranges. Specifically, we presume that the approximate analytic formulation can be extended to somewhat higher temperatures, $T\sim 100\,{\rm K}$, and somewhat lower abundances or higher velocity divergences, $\log_{_{10}}\left\{X_{_{\rm CO}}\left(|\nabla\cdot{\boldsymbol\upsilon}|/3\,{\rm km}\,{\rm s}^{-1}\,{\rm pc}^{-1}\right)^{-1}\right\}\sim -\,9$.

\section{Slow steady planar J-shocks in non-magnetic molecular clouds}\label{SEC:PSC}

As a simple example of the application of our approximate analytic formulation, we consider the role of CO in post-shock cooling layers in molecular clouds. There are many other possible applications, for example the role of CO cooling in collapsing and fragmenting pre-stellar cores, and the thermodynamics of the accretion shock bounding a growing filament. We plan to investigate these in future papers.

Molecular clouds are observed to be highly turbulent, and supersonic velocity dispersions are the norm on scales $\;\ga0.1\,{\rm pc}$ \citep{LarsonRB1981,Soloetal1987,Goodetal1998,Caseetal2002,HeyeBrun2004}. Shocks are therefore endemic in molecular clouds, and a critical issue is then how quickly the post-shock gas cools back to something like its pre-shock value. In particular, if post-shock cooling is very quick, shocks can for many purposes be treated as isothermal discontinuities, and this may greatly simplify analysis of the dynamics on larger scales. 

In the interests of simplicity, we focus on slow, steady, planar J-shocks \citep[e.g.][]{BrandPWJ1989,Flowetal2003}. We note (i) that, if the shock velocity is high, $\upsilon\subO\ga 15\,{\rm km\,s^{-1}}$, the associated kinetic energy, and resulting post-shock thermal energy, may be sufficient to drive significant chemical changes; (ii) that these chemical changes may influence the dynamical coupling between the gas, the dust and the magnetic field; and (iii) that the cooling radiation from the hot gas immediately behind the shock may have important effects on the chemistry and thermal state of the gas flowing into the shock, as will the ambient radiation field. However, these three considerations are probably not important for the low-velocity shocks, $\upsilon\subO\la 1.5\,{\rm km\,s^{-1}}$, considered here.

\subsection{The hydrodynamics of a J-shock plus cooling layer}

We consider a slow steady planar J-shock at $x\!=\!0$, with gas flowing in from $x\!<\!0$, having density $\rho(x\!<\!0)\!=\!\rho\subO$, velocity parallel to the $x$ axis $\upsilon(x\!<\!0)\!=\!\upsilon\subO$, and isothermal sound speed $a(x\!<\!0)\!=\!a\subO$. There is no magnetic field, so we are ignoring the possibility that the shock is cushioned by a lateral magnetic field; in that case, the shock would be a C-shock, and the immediate post-shock temperature would be somewhat lower, but, as shown by \citet{PonAetal2012}, the cooling would still very likely be dominated by CO. We are also ignoring the possibility that the pre-shock gas is significantly heated by the radiation from the post-shock cooling layer; this is a reasonable assumption, since gas in the post-shock cooling layer is necessarily at a very different velocity to the gas flowing into the shock, and therefore the CO cooling radiation cannot be absorbed by the inflowing gas.

We assume that the gas flowing through the shock is and remains molecular, with constant adiabatic exponent $\gamma\!=\!5/3$. This is strictly only true in low-temperature regions, $T\la 100\,{\rm K}$, where the rotational degrees of freedom of molecular hydrogen are not significantly excited \citep{WhitClar1997,Boleetal2007}. This requirement is naturally met, since the GL78 results  (and hence our approximate analytic formulation) are limited to temperatures $T\la 60\,{\rm K}$. If the pre-shock temperature is $\sim 10\,{\rm K}$, this means we must restrict our model to shocks with $\upsilon\subO\la 1.0\,{\rm km\,s^{-1}}$. If we extrapolate the approximate analytic formulation to $T\la 100\,{\rm K}$, we can treat shocks with $\upsilon\subO\la 1.5\,{\rm km\,s^{-1}}$.

The gas emerging from the shock has density, $\rho\subS$, velocity, $\upsilon\subS$, and isothermal sound speed, $a\subS$, given by the Rankine-Hugoniot conditions:
\begin{eqnarray}
\rho\subS&=&\frac{8\upsilon\subO^2\rho\subO}{(2\upsilon\subO^2+10a\subO^2)}\;\simeq\; 4\rho\subO\,;\\\label{EQN:vS}
\upsilon\subS&=&\frac{(2\upsilon\subO^2+10a\subO^2)\upsilon\subO}{8\upsilon\subO^2}\;\simeq\;\frac{\upsilon\subO}{4}\,;\\
a\subS^2&=&\frac{(2\upsilon\subO^2+10a\subO^2)(6\upsilon\subO^2-2a\subO^2)}{64\upsilon\subO^2}\;\simeq\;\frac{3\upsilon\subO^2}{16}\,,
\end{eqnarray}
where the second expressions apply provided $\upsilon\subO^2\gg a\subO^2$, and we recall that here $a\subO$ is the {\it isothermal} sound speed.

In the post-shock region ($x\!>\!0$), conservation of mass requires constant $\rho(x)\upsilon(x)$, hence
\begin{eqnarray}\label{EQN:rho(x)}
\rho(x)&=&\frac{\rho\subO \upsilon\subO}{\upsilon(x)}\,;
\end{eqnarray}
conservation of momentum requires constant $\rho(x)\left(\upsilon^2(x)+a^2(x)\right)$, hence
\begin{eqnarray}\label{EQN:MOMCON}
a^2(x)&=&\left(\upsilon\subO+\frac{a\subO^2}{\upsilon\subO}\right)\upsilon(x)-\upsilon^2(x)\,;
\end{eqnarray}
and conservation of energy requires
\begin{eqnarray}\nonumber
\frac{d}{dx}\left\{\rho(x)\upsilon(x)\left(\frac{\upsilon^2(x)}{2}+\frac{\gamma a^2(x)}{(\gamma-1)}\right)\right\}=&&\\\label{EQN:ENCON}
\rho\subO \upsilon\subO\left\{\frac{5}{2}\left(\upsilon\subO+\frac{a\subO^2}{\upsilon\subO}\right)-4\upsilon(x)\right\}\frac{d\upsilon}{dx}&=&-\Lambda\left(\rho,a,\left|\frac{d\upsilon}{dx}\right|\right),
\end{eqnarray}
where $\Lambda(\rho,a,|d\upsilon/dx|)$ is the cooling rate per unit volume, and in planar geometry $\nabla\cdot{\boldsymbol\upsilon}\rightarrow d\upsilon/dx$.

Since, in the post-shock cooling layer, $\upsilon(x)\la \upsilon\subO/4,\,$ Eqns. (\ref{EQN:MOMCON}) and (\ref{EQN:ENCON}) approximate to
\begin{eqnarray}\label{EQN:a(x)}
a^2(x)&\simeq&\upsilon\subO \upsilon(x)\,,\\\label{EQN:dvdr.1}
\frac{5\rho\subO \upsilon\subO^2}{2}\frac{d\upsilon}{dx}&\simeq&-\,\Lambda\left(\rho,a,-\frac{d\upsilon}{dx}\right),
\end{eqnarray}
where we have substituted $\;|d\upsilon/dx|=-d\upsilon/dx$, since the gas here is decelerating.

\subsection{The thermodynamics of  a J-shock plus cooling layer}\label{SEC:RES}

We assume that the abundance of CO is fixed at $X_{_{\rm CO}}\!=\!3\times 10^{-4}$. In other words, we assume that essentially all the carbon is in CO, that the abundance of C is $1.5\times 10^{-4}$ \citep[cf.][]{Sembetal2000}, and that CO is not destroyed in the shock, nor has it had time to freeze out significantly \citep[e.g.][]{Goldsmit2001}.

Substituting from Eqn. (\ref{EQN:LambdaCOTOT.1}) into Eqn. (\ref{EQN:dvdr.1}), we obtain
\begin{eqnarray}\nonumber
\frac{d\upsilon}{dx}&\simeq&-\,\frac{2\lambda_{_{\rm CO.LO}}X_{_{\rm CO}}\rho^2a^3}{5\rho\subO \upsilon\subO^2}\left\{1-\left(\!\frac{2\lambda_{_{\rm CO.HI}}a^8}{5\rho\subO \upsilon\subO^2}\!\right)^{-1/\beta}\,\right\}^{\beta}\\\label{EQN:dvdx.2}
&\simeq&-\,\frac{2\lambda_{_{\rm CO.LO}}X_{_{\rm CO}}\rho\subO \upsilon\subO^{3/2}}{5\upsilon^{1/2}}\left\{1-\left(\!\frac{2\lambda_{_{\rm CO.HI}}\upsilon\subO^2 \upsilon^4}{5\rho\subO}\!\right)^{-1/\beta}\,\right\}^{\beta}\!;
\end{eqnarray}
the second expression is obtained by substituting for $\rho$ and $a$ from Eqns. (\ref{EQN:rho(x)}) and (\ref{EQN:a(x)}). 

In Eqn. (\ref{EQN:dvdx.2}), the leading term (outside the braces) represents deceleration due to cooling in the low-density optically thin regime. The term in braces represents the correction due to the cooling trending towards the high-density optically thick regime, and this can in principle cause the post-shock cooling to stall. Since the gas must decelerate to $\upsilon\!=\!a\subO$ (see discussion following Eqn. \ref{EQN:Deltax_PSC.1} below), it follows that post-shock cooling will not stall due to the CO cooling becoming optically thick, provided that
\begin{eqnarray}\nonumber
\upsilon\subO&\gg&\left(\frac{5\,\rho\subO}{2\,\lambda_{_{\rm CO.HI}}\,a\subO^4}\right)^{\!1/2}\\
&\gg&4\times10^{-4}\,{\rm km}\,{\rm s}^{-1}\,\left(\frac{n_{_{\rm H_2.O}}}{{\rm cm}^{-3}}\right)^{\!1/2}\,\left(\frac{T\subO}{\rm K}\right)^{\!-1}\,,
\end{eqnarray}
where $n_{_{\rm H_2.O}}=\rho\subO/{\bar m}_{_{\rm H_2}}$ is the number-density of molecular hydrogen in the pre-shock gas. If this inequality is well satisfied, we can neglect the term in braces, and this is generally the case. For example, if the pre-shock density is high, say $n_{_{\rm H_2.O}}\!=\!10^5\,{\rm cm}^{-3}$, and the temperature is low, say $T\subO\!=\!10\,{\rm K}$, we require $\upsilon\subO\gg 0.012\,{\rm km}\,{\rm s}^{-1}$, which is a rather weak constraint on $\upsilon\subO$. In most cases of interest, the pre-shock density is lower than $10^5\,{\rm cm}^{-3}$, and the temperature cannot be much below $\sim 10\,{\rm K}$, in which case the minimum velocity is even lower. Furthermore, if the pre-shock density is any higher than $10^5\,{\rm cm}^{-3}$, the gas is probably so well thermally coupled to the dust that molecular line cooling is redundant \citep[e.g.][]{GlovClar2012a}.

\subsection{The thickness of the post-shock cooling layer}\label{SEC:Deltax}

The thickness of the post-shock cooling layer is
\begin{eqnarray}\label{EQN:Deltax_PSC.1}
\Delta x_{_{\rm PSC}}&=&\int\limits_{\upsilon\,\simeq\, \upsilon\subO/4}^{\upsilon\,\simeq\, a\subO}\frac{d\upsilon}{d\upsilon/dx}\,.
\end{eqnarray}
The {\it lower} limit on this integral is the immediate post-shock velocity, given by Eqn. (\ref{EQN:vS}). If we were to follow the gas until it cooled right back down to $T\subO={\bar m}\subO a\subO^2/k_{_{\rm B}}$, the {\it upper} limit on the integral would be $\upsilon\!=\!a\subO^2/\upsilon\subO$. However, by this stage the velocity would be highly subsonic and the LVG approximation would no longer be valid (in the sense that it would be seriously underestimating the cooling rate). Therefore we set the upper limit to $\upsilon\!=\!a\subO$; in other words, we follow the deceleration until the velocity becomes subsonic. We believe this is justified, since the integral in Eqn. (\ref{EQN:Deltax_PSC.1}) is dominated by the lower limit ($\upsilon=\upsilon\subO/4$).

If we substitute from Eqn. (\ref{EQN:dvdx.2}) in Eqn. (\ref{EQN:Deltax_PSC.1}), neglecting the correction term for optical thickness, we obtain 
\begin{eqnarray}\nonumber
\Delta x_{_{\rm PSC}}&\la&\frac{5}{24\,\lambda_{_{\rm CO.LO}}\,X_{_{\rm CO}}\,\rho\subO}\\\label{EQN:Deltax_PSC.3}
&\la&0.1\,{\rm pc}\,\left(\frac{X_{_{\rm CO}}}{3\!\times\!10^{-4}}\right)^{\!-1}\,\left(\frac{n_{_{\rm H_2.O}}}{{\rm cm}^{-3}}\right)^{\!-1}\,,
\end{eqnarray}
We have replaced $\simeq$ with $\la$ in Eqn. (\ref{EQN:Deltax_PSC.3}) -- and also in Eqn. (\ref{EQN:Deltat_PSC.3}) below -- because (a) we have neglected the contribution from the upper limit in the integral of Eqn. (\ref{EQN:Deltax_PSC.1}), and this contribution can be quite large for relatively slow shocks, thereby reducing $\Delta x_{_{\rm PSC}}$ further; (b) for fast shocks there will be a significant additional cooling contribution from dust \citep{GlovClar2012a}, and possibly also other molecules \citep{Neufetal1995}, due to the high post-shock density, reducing $\Delta x_{_{\rm PSC}}$ still further.

The gas passes through $\Delta x_{_{\rm PSC}}$ in a time
\begin{eqnarray}\label{EQN:Deltat_PSC.1}
\Delta t_{_{\rm PSC}}&=&\int\limits_{\upsilon\,\simeq\, \upsilon\subO/4}^{\upsilon\,\simeq\, a\subO}\frac{d\upsilon}{\upsilon\;d\upsilon/dx}\\\nonumber
&\la&\frac{5}{2\lambda_{_{\rm CO.LO}}X_{_{\rm CO}}\rho\subO \upsilon\subO}\\\label{EQN:Deltat_PSC.3}
&\la&1.2\,{\rm Myr}\;\left(\frac{X_{_{\rm CO}}}{3\!\times\!10^{-4}}\right)^{\!-1}\!\left(\frac{n_{_{\rm H_2.O}}}{{\rm cm}^{-3}}\right)^{\!-1}\!\left(\frac{\upsilon\subO}{{\rm km}\,{\rm s}^{\!-1}}\right)^{\!-1}\!;
\end{eqnarray}
and the integrated flux in CO cooling lines from the shock is
\begin{eqnarray}
F_{_{\rm CO}}\!&\!\la\!&\!\frac{5\rho\subO \upsilon\subO^3}{8}\\
\!&\!\la\!&\!3\!\times\! 10^{-9}\,{\rm erg}\,{\rm cm}^{-2}\,{\rm s}^{-1}\;\left(\frac{n_{_{\rm H_2.O}}}{{\rm cm}^{-3}}\right)\,\left(\frac{\upsilon\subO}{{\rm km}\,{\rm s}^{\!-1}}\right)^3.
\end{eqnarray}

\subsection{Caveats on shock model}\label{SEC:Caveats}

As already stated, gas-phase CO appears to be the dominant form of carbon at volume densities in the range $3\times 10^2\,{\rm H}_2\,{\rm cm}^{-3}\la n_{_{\rm H_2}}\la 3\times 10^5\,{\rm H}_2\,{\rm cm}^{-3}$, and therefore our shock model can only be applied to shocks involving pre-shock gas with density, $n_{_{\rm H_2.O}}$, in this range. Many shocks arising in turbulent molecular clouds will involve gas with density in this range.

J-shocks in turbulent molecular clouds will only approximate to being steady and planar if the counter-flows that form them have coherence lengths significantly larger than the thickness of the post-shock cooling layer, $\Delta x_{_{\rm PSC}}$. If we adopt coherence lengths from Larson's relations, i.e. $L\sim 1600\,{\rm pc}\,(n_{_{\rm H_2.O}}/{\rm cm^{-3}})^{-0.91}$ \citep{LarsonRB1981}, this condition is very well satisfied. For example, if we set $X_{_{\rm CO}}=3\times 10^{-4}$, we have $\Delta x_{_{\rm PSC}}/L\sim 6\times 10^{-5}\,(n_{_{\rm H_2}}/{\rm cm}^{-3})^{-0.09}$. This suggests that the assumption of a steady planar shock is reasonable.

If we assume that the approximate analytic formulation can be extended to $T\sim 100\,{\rm K}$, and that the pre-shock gas has $T\sim 10\,{\rm K}$, then our shock model can only be applied to shocks with  pre-shock velocity $\upsilon\subO\la 1.5\,{\rm km\,s^{-1}}$, otherwise the immediate post-shock gas-kinetic temperature (i.e. pre cooling) is too high. This is a significant, but not critical, restriction, since there are likely to be many shocks satisfying this constraint.

For example, column-density maps of low-mass cores \citep{Konyetal2015} and filaments \citep{Palmetal2013} have centrally condensed profiles. However, the column-density contrast between the background and the line of sight through the centre of the core or the spine of the filament is small, typically $\la 10$. This indicates that the ram pressure of the gas accreting onto the core or filament is low, otherwise the profile would be flat. It also suggests that the background volume-density is not hugely lower than that inside the core or filament, hence the Mach number of the approximately isothermal accretion shock at the boundary of the core or filament must be low, ${\cal M}\la 5$, and the pre-shock velocity must also be low, $\upsilon\subO\la 1\,{\rm km\,s^{-1}}$. In addition we note that gravitational acceleration of the material accreting onto a marginally Jeans-unstable core or Ostriker-unstable filament can only generate marginally trans-sonic velocities, so gravitational acceleration does not alter this conclusion.

To first order, we can approximate 
\begin{eqnarray}
|\nabla\cdot{\boldsymbol\upsilon}|\sim\frac{\upsilon\subO}{\Delta x_{_{\rm PSC}}}\;\,\simeq\;\,\frac{\upsilon\subO}{0.1\,{\rm pc}}\,\left(\frac{X_{_{\rm CO}}}{3\!\times\!10^{-4}}\right)\,\left(\frac{n_{_{\rm H_2.O}}}{{\rm cm}^{-3}}\right)\,,
\end{eqnarray}
using Eqn. (\ref{EQN:Deltax_PSC.3}). If again we set $X_{_{\rm CO}}=3\times 10^{-4}$, then the limits on $X_{_{\rm CO}}^{-1}|\nabla\cdot{\boldsymbol\upsilon}|$, i.e. $-9\la \log_{_{10}}\left\{X_{_{\rm CO}}\left(|\nabla\cdot{\boldsymbol\upsilon}|/3\,{\rm km}\,{\rm s}^{-1}\,{\rm pc}^{-1}\right)^{-1}\right\}\la -4$ yield
\begin{eqnarray}
2.4&\la&\left(\frac{\upsilon\subO}{\rm km\,s^{-1}}\right)\,\left(\frac{n_{_{\rm H_2.O}}}{{\rm cm}^{-3}}\right)\;\,\la\;\,2.4\times10^5\,.
\end{eqnarray}
Since we already have the constraint  $\upsilon\subO \la 1.5\,{\rm km\,s^{-1}}$, this reduces to $n_{_{\rm H_2.O}}\la 1.6\times 10^5\,{\rm cm}^{-3}$, which is only slightly more restrictive than the upper limit on $n_{_{\rm H_2}}$ derived at the start of this section.

Our shock model assumes that the carbon in the pre-shock gas is mainly in CO, and that the CO is not destroyed in the shock, nor does it freeze out. If this is not the case, cooling will be provided by other species, for example C$^+$ \citep[see][]{GlovClar2012b}, but the gas will not cool so fast, and will probably not be able to get down to $\sim 10\,{\rm K}$, unless the post-shock density becomes high enough for the gas to couple thermally to the dust.

\subsection{Discussion of shock results}\label{SEC:DISC}

The thickness of the post-shock cooling layer, $\Delta x_{_{\rm PSC}}$, and the post-shock cooling time, $\Delta t_{_{\rm PSC}}$, are both likely to be very small, as compared with other length- and time-scales in the molecular cloud. With typical pre-shock densities in a molecular cloud, $n_{_{\rm H_2.O}}\ga 3\times 10^2\,{\rm cm}^{-3}$, and setting $X_{_{\rm CO}}\simeq 3\!\times\!10^{-4}$, we have $\Delta x_{_{\rm PSC}}\,\la\, 0.0003\,{\rm pc}\;(\equiv\!\! 70\,{\rm AU})$ and $\Delta t_{_{\rm PSC}}\,\la\, 4\,{\rm kyr}\,(\upsilon\subO/{\rm km}\,{\rm s}^{-1})^{-1}$. These values are similar to the values derived by \citet{PonAetal2014,PonAetal2016a}, although they treated magnetically cushioned C-shocks, in which the compression and heating are less abrupt and less extreme. 

One reason why CO is an effective post-shock coolant is that it has a low dipole moment, and therefore its lines remain optically thin up to quite high column-densities. A second reason is that the steady deceleration of the post-shock gas ensures that there is a broad range of velocities, and hence wavelengths, over which each line can be emitted, and therefore the lines do not readily become optically thick.

When treating the dynamics of turbulent interstellar clouds -- for example, their assembly, cloud/cloud collisions, the formation of sheets and filaments and cores -- Eqns.  (\ref{EQN:Deltax_PSC.3}) and (\ref{EQN:Deltat_PSC.3}) can be used to assess rather quickly whether it is acceptable to treat the associated shocks as isothermal discontinuities, or alternatively, what resolution is needed to resolve the post-shock cooling layers. That is, provided that the inherent assumptions in our analysis are valid, viz. steady slow non-magnetic planar J-shocks, in which CO dominates the cooling and is not destroyed or removed from the gas-phase (as discussed in Section \ref{SEC:Caveats}).

\citet{PonAetal2012} and \citet{PonAetal2016a} estimate that the volume fraction of a turbulent molecular cloud that is involved in shock dissipation (i.e. the fraction that is in post-shock cooling layers) is $f_{_{\rm PSC}}\sim 0.001$. Given the short cooling times estimated above (Eqn. \ref{EQN:Deltat_PSC.3}), it follows that a representative fluid element in such a cloud should be shocked at least once every Myr. Indeed, \citet{PanLPado2009} identify such shocks as an important overall heating mechanism for molecular clouds, although our estimates indicate that this heating should be very localised and transient.

\section{Conclusions}\label{SEC:CONC}

We have derived a simple approximate analytic formulation for the cooling rate due to rotational transitions of the CO molecule,
\begin{eqnarray}\left.\begin{array}{rcl}
\Lambda_{_{\rm CO.LO}}&\simeq&\left[4.63\times 10^8\,{\rm cm}^2\,{\rm g}^{-1}\right]\,X_{_{\rm CO}}\,\rho^2a^3\,,\\&&\\
\Lambda_{_{\rm CO.HI}}&\simeq&\left[4.58\times 10^{-45}\,{\rm cm}^{-9}\,{\rm g}\,{\rm s}^6\right]\,a^8\,|\nabla\cdot{\boldsymbol\upsilon}|\,,\\&&\\
\beta&\simeq&0.813\,\left(\frac{\rho}{{\rm g}\,{\rm cm}^{-3}}\right)^{0.0533}\,\left(\frac{a}{{\rm cm}\,{\rm s}^{-1}}\right)^{0.328}\,,\\&&\\
\Lambda_{_{\rm CO.TOT}}&\simeq&\left\{\Lambda_{_{\rm CO.LO}}^{-1/\beta}+\Lambda_{_{\rm CO.HI}}^{-1/\beta}\right\}^{-\beta}\,.
\end{array}\right\}\end{eqnarray}
An equivalent formulation, but giving $\Lambda_{_{\rm CO.TOT}}/n_{_{\rm CO}}$ in terms of $n_{_{\rm H_2}}$ and $T$ (instead of $\rho$ and $a$), is given in Eqns. (\ref{EQN:Lambda_COLO.2}) through (\ref{EQN:Lambda_COTOT.2}). Our approximate analytic formulation extends from the low-density optically thin limit, to the high-density optically thick limit, and includes the contributions from the different isotopologues of CO. It is based on physical considerations, but is calibrated against the detailed numerical results of GL78, and reproduces those results well (see Fig. \ref{FIG:GoldLangFig2}). It should be usable for
\begin{eqnarray}\nonumber
10\,{\rm K}\;\,\la\;\,T\;\,\la\;\,100\,{\rm K}\,,\hspace{2.5cm}\\\nonumber
3\times 10^2\,{\rm cm}^{-3}\;\,\la\;\,n_{_{\rm H_2}}\;\,\la\;\,3\times 10^5\,{\rm cm}^{-3}\,,\hspace{1.4cm}\\\nonumber
10\,{\rm km\,s^{-1}\,pc^{-1}}\;\,\la\;\,|\nabla\cdot{\boldsymbol\upsilon}|\,\left(\frac{X_{_{\rm CO}}}{3\times 10^{-4}}\right)^{-1}\;\,\la\;\,10^6\,{\rm km\,s^{-1}\,pc^{-1}}\,.
\end{eqnarray} 

Using this formulation, we have derived estimates of the thickness of post-shock cooling layers, $\Delta x_{_{\rm PSC}}$, and post-shock cooling times, $\Delta t_{_{\rm PSC}}$, under the assumptions (i) that the shocks are steady slow non-magnetic planar J-shocks, and (ii) that CO dominates the cooling, and is not destroyed or removed from the gas-phase in the shock or the post-shock cooling layer. With these caveats, and given the constraints on shock velocity, pre-shock density and pre-shock temperature detailed in Section \ref{SEC:Caveats}, our estimates can be used to justify treating shocks as isothermal discontinuities when the primary concern is dynamics on larger scales; to evaluate the resolution required to model a shock; and to estimate the integrated CO flux from a post-shock cooling region. The values of $\Delta t_{_{\rm PSC}}$ suggest that a typical fluid element in a molecular cloud should be shocked at least once every Myr.

\begin{acknowledgements}
We thank the anonymous referee for a very thoughtful and constructive report, which greatly improved the original version of this paper. APW gratefully acknowledges the support of a consolidated grant (ST/K00926/1), and SEJ a PhD studentship, both from the UK Science and Technology Funding Council. This work was performed using the computational facilities of the Advanced Research Computing at Cardiff (ARCCA) Division, Cardiff University. 
\end{acknowledgements}

\bibliographystyle{aa}
\bibliography{Whitworth}

\end{document}